\documentclass[twocolumn,prl,showpacs,preprintnumbers,amsmath,amssymb]{revtex4}
\usepackage{graphicx}
\usepackage{dcolumn}
\usepackage{bm}

\begin{document}

\title{Spin-Hall Conductivity in Electron-Phonon Coupled Systems}
\author{C. Grimaldi,$^{1,2}$ E. Cappelluti,$^{3,4}$ and F. Marsiglio$^{2,5}$}

\affiliation{$^1$ LPM, Ecole Polytechnique F\'ed\'erale de
Lausanne, Station 17, CH-1015 Lausanne, Switzerland}

\affiliation{$^2$DPMC, Universit\'e de Gen\`{e}ve, 24 Quai
Ernest-Ansermet, CH-1211 Gen\`{e}ve 4, Switzerland}

\affiliation{$^3$Istituto dei Sistemi Complessi, CNR-INFM, via dei
Taurini 19, 00185 Roma, Italy}

\affiliation{$^4$Dipartimento di Fisica, Universit\`a ``La
Sapienza'', Piazzale Aldo Moro 2, 00185 Roma, Italy}

\affiliation{$^5$Department of Physics, University of Alberta,
Edmonton, Alberta, Canada, T6G 2J1}

\begin{abstract}
We derive the ac spin-Hall conductivity $\sigma_{\rm sH}(\omega)$
of two-dimensional spin-orbit coupled systems interacting with
dispersionless phonons of frequency $\omega_0$. For the linear
Rashba model we show that the electron-phonon contribution to the
spin-vertex corrections breaks the universality of $\sigma_{\rm
sH}(\omega)$ at low-frequencies and provides a non-trivial
renormalization of the interband resonance. On the contrary, in a
generalized Rashba model for which the spin-vertex contributions
are absent, the coupling to the phonons enters only through the
self-energy, leaving the low frequency behavior of $\sigma_{\rm
sH}(\omega)$ unaffected by the electron-phonon interaction.
\end{abstract}

\maketitle

The recent prediction of intrinsic spin currents generated by
applied electric fields in semiconductors with spin-orbit (SO)
interaction \cite{zhang,sinova} has attracted intensive research
on the subject \cite{schlie,engel} encouraged also by potential
applications in spintronic-based devices. In such systems, the
spin-Hall conductivity $\sigma_{\rm sH}=J_y^{S_z}/E_x$, where
$J_y^{S_z}$ is a spin $S_z$ polarized current in the $y$-direction
and $E_x$ is the electric field directed along $x$, arises from
the SO dependent band structure which, for clean systems, leads
for example to $\sigma_{\rm sH}=-e/8\pi$ for a two-dimensional
(2D) electron system with Rashba SO coupling \cite{sinova} or to
$\sigma_{\rm sH}=-3e/8\pi$ for a 2D hole semiconductor
\cite{loss2,bernevig}.

Of special interest for both applied and fundamental research is
the role played by scattering events which have been shown to
modify in an essential way the clean limit results. The most
drastic effects are found in the 2D linear Rashba model, where
$\sigma_{\rm sH}$ reduces to zero for arbitrarily weak impurity
scattering \cite{inoue,halpe,raimo,loss,dimi}, while the universal
value $\sigma_{\rm sH}(\omega)=-e/8\pi$ is recovered for finite
values of the ac field frequency $\omega$ in the range $\tau^{-1}<
\omega < \Delta$ \cite{halpe,loss}, where $\tau^{-1}$ is the
impurity scattering rate and $\Delta$ is the spin-orbit energy
splitting. On the contrary, in 2D hole systems with weak (short
ranged) impurity scattering, $\sigma_{\rm sH}(\omega)$ remains
equal to $-3e/8\pi$ for $0\leq \omega < \Delta$
\cite{loss2,bernevig,nomura}, while it becomes dependent on the
impurity potential if this has long range character
\cite{shytov,khae}.

So far, the study of scattering effects on the spin-Hall
conductivity has been restricted to the case in which the source
of scattering is the coupling of the charge carriers to some
elastic impurity potential. This leaves aside the contributions
from inelastic scattering such those provided by the
electron-phonon (el-ph) interaction which, in the materials of
interest for the spin-Hall effect, ranges from the weak-coupling
limit in GaAs \cite{dassarmaold} to the strong-coupling regime in
Bi(100) \cite{gayone}.

Because of its dynamic and inelastic character, the el-ph
interaction may affect the spin-Hall response in a way
drastically different from static elastic impurity scattering,
questioning the general validity of the commonly accepted forms
of $\sigma_{\rm sH}(\omega)$ summarized above. Furthermore, the
issue of the vertex corrections, which are responsible for the
vanishing of $\sigma_{\rm sH}(\omega=0)$ in the impure 2D linear
Rashba model \cite{inoue,raimo,loss,dimi,dassarma}, acquires a
new importance, since these should be altered by the el-ph
interaction.

In this letter we report on our results on the spin-Hall
conductivity $\sigma_{\rm sH}(\omega)$ for 2D systems with SO
interaction coupled with dispersionless phonons of frequency
$\omega_0$. For a linear Rashba model we show that, in the
frequency range $\tau^{-1} < \omega \ll \Delta$ (with $\Delta
<\omega_0$) where the universal value $-e/8\pi$ has been
predicted, the el-ph contribution to the vertex corrections
reduces $\sigma_{\rm sH}(\omega)$ to the nonuniversal value
$-e/[8\pi(1+\lambda/2)]$, where $\lambda$ is the el-ph coupling
constant. Furthermore, we find that the el-ph spin-vertex
contributions renormalize also the interband transitions and
provide a further reduction of $\sigma_{\rm sH}(\omega)$ for
$\omega>\omega_0$. On the contrary, in a 2D generalized Rashba
model, for which the spin-vertex contributions are absent, the
el-ph interaction provides only a trivial self-energy correction
to the interband transition, leaving the low frequency part of
$\sigma_{\rm sH}(\omega)$ basically unaltered.

We consider the el-ph interaction as given by the Holstein
hamiltonian generalized to include SO coupling:
\begin{eqnarray}
\label{h1} H&=&\sum_{{\bf k},\alpha}\epsilon_{\bf k}
c^\dagger_{{\bf k}\alpha}c_{{\bf k}\alpha}+\sum_{{\bf
k}\alpha\beta}\mathbf{\Omega}_{\mathbf{k}}
\cdot\mbox{\boldmath$\sigma$}_{\alpha\beta}c^\dagger_{{\bf
k}\alpha}c_{{\bf k}\beta} \nonumber \\
&&+\omega_0\sum_{\bf q} a^\dagger_{\bf q}a_{\bf q} +g\sum_{{\bf
q}{\bf k}\alpha}c^\dagger_{{\bf k}\alpha}c_{{\bf k}-{\bf
q}\alpha}(a_{\bf q}+a^\dagger_{-{\bf q}}),
\end{eqnarray}
where $c^\dagger_{{\bf k}\alpha}$ and $a^\dagger_{\bf q}$
($c_{{\bf k}\alpha}$ and $a_{\bf q}$) are the creation
(annihilation) operators for an electron with momentum ${\bf
k}=(k_x,k_y)$ and spin index $\alpha=\uparrow,\downarrow$, and for
a phonon with wavenumber ${\bf q}$. $\epsilon_{\bf
k}=\hbar^2k^2/2m$ is the electron dispersion, $\omega_0$ is the
phonon frequency and $g$ is a momentum independent  el-ph
interaction (Holstein model). The use of a Holstein coupling
permits to focus solely on the retardation and inelastic effects
of phonons, disentangling the study from possible momentum
dependences of the el-ph interaction. Furthermore, the Holstein
coupling is partially justified, for example, by the results on
surface states \cite{gayone} and by the reduced momentum
dependence, compared to 3D electron gases, of 2D electrons
coupled to bulk polar optical phonons \cite{dassarmaold}. In the
following, we shall also include the coupling to a short-ranged
impurity potential $V({\bf r})=V_{\rm imp}\sum_i\delta({\bf
r}-{\bf R}_i)$, where ${\bf R}_i$ are the random positions of the
impurity scatterers.

Let us start by considering a linear Rashba model, for which the
SO vector potential is $\mathbf{\Omega}_{\mathbf{k}}=\gamma k
(-\sin\phi,\cos\phi)$, where $\gamma$ is the SO coupling and
$\phi$ is the polar angle. The electron Green's function of the
interacting system is
\begin{equation}
\label{green1} G({\bf k},i\omega_n)=\frac{1}{2}\sum_{s=\pm
1}[1+s\hat{\mathbf{\Omega}}_{\mathbf{k}}\cdot\mbox{\boldmath$\sigma$}]
G_s(k,i\omega_n),
\end{equation}
where $\hat{\mathbf{\Omega}}_{\mathbf{k}}=(-\cos\phi,\sin\phi)$
and
$G_s(k,i\omega_n)=[i\omega_n-E_k^s+\mu-\Sigma(i\omega_n)]^{-1}$
is the Green's function in the helicity basis with dispersion
$E_k^s=\hbar^2(k+sk_0)^2/2m$, $k_0=m\gamma/\hbar^2$ is the SO
wavenumber, $\mu$ is the chemical potential and
$\omega_n=(2n+1)\pi T$ is the fermionic Matsubara frequency at
temperature $T$. Due to the momentum independence of $g$ and
$V_{\rm imp}$, the self-energy $\Sigma(i\omega_n)$ is independent
of ${\bf k}$ and reduces to
\begin{equation}
\label{self1}
\Sigma(i\omega_n)=T\!\sum_{n'}\frac{W(i\omega_n-i\omega_{n'})} {2
N_0}\sum_{s=\pm }\int\!\frac{dk}{2\pi} k \, G_s(k,i\omega_{n'}),
\end{equation}
where $N_0=m/2\pi\hbar^2$ is the density of states per spin
direction and
\begin{equation}
\label{self2}
W(i\omega_n-i\omega_{n'})=\frac{\delta_{n,n'}}{2\pi\tau
T}-\lambda\frac{\omega_0^2}{(i\omega_n-i\omega_{n'})-\omega_0^2},
\end{equation}
where $\tau^{-1}=2\pi n_iV^2_{\rm imp}N_0$ is the impurity
scattering rate and $\lambda=2 g^2 N_0/\omega_0$ is the el-ph
coupling. In writing Eqs.(\ref{self1},\ref{self2}), we have
employed the self-consistent Born approximation for both impurity
and el-ph scatterings.

The equations defining the spin-Hall conductivity are obtained
from the Kubo formula applied to the el-ph problem. Hence, the
spin-current--charge-current correlation function is
\begin{equation}\label{sh3}
K(i\nu_m)=i\frac{e\hbar^2\gamma}{4m}T\sum_n\Gamma(i\omega_l,i\omega_n)
B_1(i\omega_l,i\omega_n),
\end{equation}
where $\nu_m=2m\pi T$ is a bosonic Matsubara frequency,
$\omega_l=\omega_n+\nu_m$ and
\begin{equation}
\label{B1} B_1(i\omega_l,i\omega_n)=\int\!\frac{dk}{2\pi}
k^2\!\sum_s\!s\,G_{-s}(k,i\omega_l)G_s(k,i\omega_n).
\end{equation}
The vertex function $\Gamma$ appearing in Eq.(\ref{sh3})
satisfies the following self-consistent equation
\begin{eqnarray}
\label{gamma1} \Gamma(i\omega_l,i\omega_n)&=&1+
T\sum_{n'}\frac{W(i\omega_{n'}-i\omega_n)}{4N_0k_0}
\left[B_2(i\omega_{l'},i\omega_{n'})\right.
\nonumber\\
&&\left.+k_0B_3(i\omega_{l'},i\omega_{n'})
\Gamma(i\omega_{l'},i\omega_{n'})\right],
\end{eqnarray}
where $\omega_{l'}=\omega_{n'}+\nu_m$ and
\begin{eqnarray}
\label{bs2}B_2(i\omega_l,i\omega_n)&=&\int\!\frac{dk}{2\pi}
k^2\!\sum_s\!s\,G_{s}(k,i\omega_l)G_s(k,i\omega_n),\;\;\;\;\;\; \\
\label{bs3}B_3(i\omega_l,i\omega_n)&=&\int\!\frac{dk}{2\pi}
k\!\sum_{s,s'}\,G_{s}(k,i\omega_l)G_{s'}(k,i\omega_n).\;\;\;\;\;\;
\end{eqnarray}
All integrations over the momenta $k$ appearing in the above
equations can be performed analytically, while the
self-consistent equations (\ref{self1}) and  (\ref{gamma1}) are
solved numerically by iteration in the Matsubara frequency space.
Finally, the (complex) spin-Hall conductivity
\begin{equation}
\label{sigma} \sigma_{\rm sH}(\omega)=i\frac{K^R(\omega)}{\omega}
\end{equation}
is obtained from the retarded function $K^R(\omega)\equiv
K(\omega+i\delta)$ extracted from $K(i\nu_m)$ [Eq.(\ref{sh3})] by
applying the Pad\'e method of numerical analytical continuation.
Although our numerical calculations can be applied to arbitrary
values of $\Delta/E_F$, where $E_F$ is the Fermi energy and
$\Delta=2\gamma k_F$ is the SO splitting, the following
discussion will be restricted to the weak SO coupling limit
$\Delta/E_F\ll 1$, common to many materials, for which some
analytical results can be obtained. In our calculations we have
used $T=0.01\omega_0$ (or $T=0.001\omega_0$ for the case shown in
Fig. \ref{fig2}), which is representative of the zero temperature
case.

\begin{figure}[t]
\protect
\includegraphics[scale=0.4]{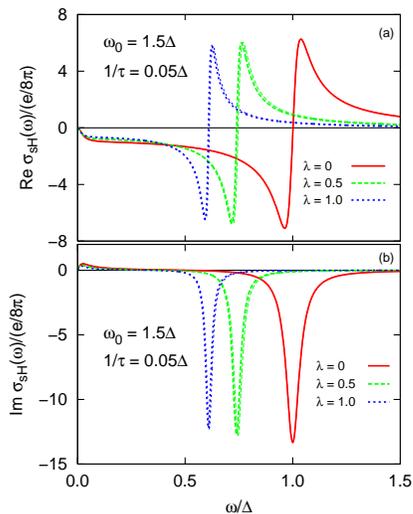}
\caption{(color online) Real (a) and imaginary (b) parts of the
spin-Hall conductivity for $\omega_0>\Delta$ obtained from the
numerical analytical continuation of
Eqs.(\ref{sh3})-(\ref{sigma}) (thick lines). The analytical
formula (\ref{sigma2}) is plotted with thin lines and is almost
indistinguishable from the numerical results. The peaks of ${\rm
Im}\,\sigma_{\rm sH}(\omega)$ are centered at
$\omega/\Delta=\sqrt{1+\lambda/2}/(1+\lambda)$.} \label{fig1}
\end{figure}

We start our analysis by considering first the case $\omega_0
>\Delta$ for which, as discussed below, the el-ph effects enter
mainly through the real parts of the self-energy and of the
vertex function. In Fig. \ref{fig1} we show the real and
imaginary parts of the spin-Hall conductivity for
$\omega_0=1.5\Delta$ and $\lambda=0,\,0.5,\,1.0$, and for weak
impurity scattering $1/\tau\Delta=0.05$. In the absence of el-ph
interaction ($\lambda=0$), we recover the known results
\cite{halpe,loss} characterized by the strong interband
transitions at $\omega=\Delta$ and by the vanishing of
$\sigma_{\rm sH}(\omega)$ as $\omega\rightarrow 0$. Furthermore,
in the intermediate-frequency region $1/\tau<\omega\ll \Delta$,
$\mathrm{Re}\, \sigma_{\rm sH}(\omega)$ is almost
$\omega$-independent and matches the universal value $-e/8\pi$.
This is better displayed in Fig. \ref{fig2}(a) where the
low-frequency behavior is plotted for $1/\tau\Delta=0.005$. Upon
enhancing $\lambda$, two new features emerge. Namely, the
frequency of the interband transitions get shifted at a lower
($\lambda$ dependent) value and, as also shown in Fig.
\ref{fig2}(a), the intermediate-frequency real spin-Hall
conductivity deviates from $-e/8\pi$, indicating that
universality breaks down when $\lambda\neq 0$.  The origin of
these features can be understood from the analysis of Eqs.
(\ref{sh3}) and (\ref{gamma1}). In fact, at zero temperature and
for $\Delta/E_F\ll 1$, the retarded function $K^R(\omega)$
reduces to \cite{mahan}
\begin{equation}\label{sh3b}
K^R(\omega)=\frac{e\hbar^2\gamma}{4m}\int_{-\omega}^{0}\!\frac{d\epsilon}{2\pi}
\Gamma(\epsilon_+ +\omega,\epsilon_-) B_1(\epsilon_+
+\omega,\epsilon_-),
\end{equation}
where $\epsilon_\pm=\epsilon\pm i\delta$. For $\omega<\omega_0$,
the integration appearing in (\ref{sh3b}) restricts the
$\epsilon+\omega$ and $\epsilon$ variables to
$|\epsilon+\omega|<\omega_0$ and $|\epsilon|<\omega_0$, for which
the self-energy on the real axis can be well approximated by
$\Sigma(x_\pm) = -\lambda {\rm Re}( x_\pm)\mp i/2\tau$, where
$x_-=\epsilon_-$ and $x_+=\epsilon_++\omega$. In this way, the
quite lengthy integral equation for $\Gamma(\epsilon_+
+\omega,\epsilon_-)$, which can be derived from Eq.(\ref{gamma1})
by following the method of analytic continuation described in
Ref.\cite{mahan}, reduces to a simple $\omega$-dependent
algebraic equation. Its solution for $\omega/\Delta\ll 1$ is
$\Gamma(\omega)\simeq \omega/[(1+\lambda/2)\omega+i/2\tau]$ and,
since $B_1(\omega)$ is a constant for $\omega/\Delta\ll 1$, the
low-frequency spin-Hall conductivity becomes $\sigma_{\rm
sH}(\omega)\simeq -(e/8\pi)\Gamma(\omega)$. We recover therefore
the vanishing of $\sigma_{\rm sH}(\omega)$ for $\omega\rightarrow
0$ while, contrary to the $\lambda=0$ case, we find that
$\sigma_{\rm sH}(\omega)$ is approximately equal to the
nonuniversal constant $-e/[8\pi(1+\lambda/2)]$ for
$1/\tau<\omega\ll \Delta$. The breakdown of universality at
intermediate frequency reported in Figs. \ref{fig1}(a) and
\ref{fig2}(a) stems therefore from the el-ph contribution to the
spin-vertex correction which, from Eqs.(\ref{gamma1}-\ref{bs3}),
governs the intraband contributions to $\sigma_{\rm sH}(\omega)$.
A more refined calculation which takes into account also the
interband transitions leads to:
\begin{equation}
\label{sigma2} \sigma_{\rm
sH}(\omega)=-\frac{e}{8\pi}\frac{\omega}
{\left(1+\frac{\lambda}{2}\right)\omega+\frac{i}{2\tau}-
\left[(1+\lambda)\frac{\omega}{\Delta}+\frac{i}{\tau\Delta}\right]^2\omega},
\end{equation}
which is valid for $\omega<\omega_0$ and arbitrary
$\omega/\Delta$ (for $\Delta/E_F\ll 1$). For $\lambda=0$,
Eq.(\ref{sigma2}) is identical to the formula already published
in Refs.\cite{halpe,loss}. Instead, for $\lambda>0$ we recover
the intermediate frequency nonuniversal behavior discussed above
together with an el-ph renormalization effect to the interband
transitions, which now occur at a frequency $\omega=\Delta^*$,
where
\begin{equation}
\label{reno} \Delta^*=\frac{\sqrt{1+\lambda/2}}{1+\lambda}\Delta
\end{equation}
for $1/\tau\Delta\ll 1$. When compared with the numerical results
of Figs. \ref{fig1} and \ref{fig2}(a), equation (\ref{sigma2}) is
in excellent agreement for all frequencies lower than $\omega_0$.
As a matter of fact, Eq.(\ref{sigma2}) is in very good agreement
with the numerical results also for $\omega>\omega_0$ as long as
$\omega_0 > \Delta$ while, for $\omega_0 <\Delta$, the
$\omega$-dependence of $\sigma_{\rm sH}(\omega)$ starts to be
affected by the imaginary contributions of the el-ph self-energy
and of the vertex function. These effects  are visible in Fig.
\ref{fig2}(b), where we compare the numerical results for
$\omega_0=0.05 \Delta$ (thick lines) with Eq.(\ref{sigma2}) (thin
lines). The deviation of ${\rm Re}\,\sigma_{\rm sH}(\omega)$ from
$-e/[8\pi(1+\lambda/2)]$ for $\omega\gtrsim\omega_0$ stems from
intraband transitions mediated by the phonons which, in analogy
to the low temperature optical conductivity of the Holstein el-ph
model \cite{bickers,marsiglio97}, ensure conservation of energy
and momentum. At higher frequencies, the real part of the el-ph
self-energy goes to zero as $\lambda\omega_0^2/\omega$ for large
$\omega/\omega_0$, and the interband transitions occur at the
unrenormalized frequency $\omega\approx\Delta$.

\begin{figure}[t]
\protect
\includegraphics[scale=0.4]{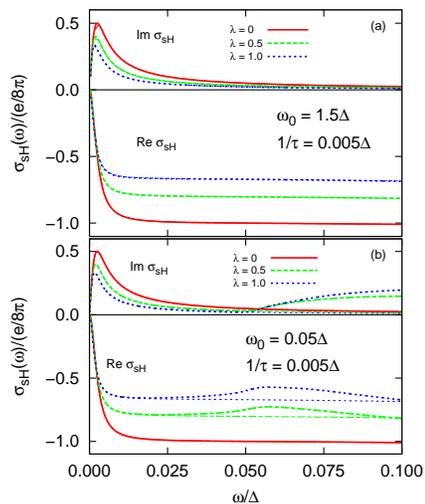}
\caption{(color online) Low frequency behavior of $\sigma_{\rm
sH}(\omega)$ for $\omega_0>\Delta$ (a) and for $\omega_0\ll
\Delta$ (b). The thick lines are the numerical results, while the
thin lines are Eq.(\ref{sigma2}). In panel (a) they are barely
distinguishable.} \label{fig2}
\end{figure}

\begin{figure}[t]
\protect
\includegraphics[scale=0.3]{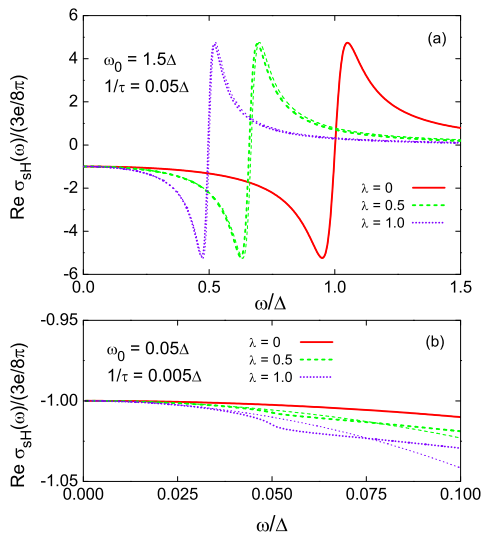}
\caption{(color online) (a): Real part of the spin-Hall
conductivity for a generalized Rashba model with $N=3$. Thick
lines are the numerical results while thin lines are
Eq.(\ref{shhole}). (b): low-frequency region for $\omega_0\ll
\Delta$.} \label{fig3}
\end{figure}

Having established that the nonuniversality of $\sigma_{\rm
sH}(\omega)$ at intermediate frequencies and the non-trivial
renormalization (\ref{reno})  have their origin in the el-ph
contributions to the spin-vertex correction, we now turn to
evaluate the el-ph effects when the spin-vertex corrections are
absent. To investigate this point we have considered a 2D
generalized Rashba model where the SO interaction is of the form
$\mathbf{\Omega}_{\mathbf{k}}=\gamma k^N (-\sin N\phi,\cos
N\phi)$ \cite{khae}. For $N=1$, the linear Rashba model discussed
above is recovered, while for $N=3$ this model describes a 2D
hole gas subjected to an asymmetric confining potential. Because
of the angular dependence of $\mathbf{\Omega}_{\mathbf{k}}$ for
$N\neq 1$, the vertex corrections are absent \cite{bernevig}, and
the correlation function $K(i\nu_m)$ is simply given by
Eq.(\ref{sh3}) with $\Gamma(i\omega_l,i\omega_n)=1$ and with the
prefactor multiplied by $N$. Furthermore, the function
$B_1(i\omega_l,i\omega_n)$ is as given in Eq.(\ref{B1}), with $dk
k^2$ replaced by $dk k^{1+N}$ and with dispersion $E_k^s=\hbar^2k
^2/2m+s\gamma k^3$. Contrary to the linear Rashba model, now all
el-ph effects arise solely from the self-energies contained in
the interband bubble  term $B_1$. Hence, in the weak SO limit
$\Delta/E_F\ll 1$, where now $\Delta=2\gamma k_F^3$, and by using
the same approximation scheme as above, for $\Delta,\omega
<\omega_0$  the spin-Hall conductivity is easily found to be
given by:
\begin{equation}
\label{shhole} \sigma_{\rm
sH}(\omega)=-\frac{eN}{8\pi}\frac{\Delta^2}
{\Delta^2-[(1+\lambda)\omega+i/\tau]^2}.
\end{equation}
Contrary to Eq.(\ref{sigma2}), the above expression predicts a
low-frequency behavior unaffected by the el-ph interaction.
Namely: $\sigma_{\rm sH}(\omega)=-eN/8\pi$ for $\omega\ll\Delta$.
Furthermore, the interband transition frequency is renormalized
only by the el-ph self-energy (mass enhancement) factor
$1+\lambda$: $\Delta^*=\Delta/(1+\lambda)$, in contrast with
Eq.(\ref{reno}) where the el-ph contribution to the spin-vertex
corrections contributes with a factor $\sqrt{1+\lambda/2}$. This
behavior is confirmed by our numerical results for $N=3$ reported
in Fig. \ref{fig3}(a) (thick lines), which fully agree with Eq.
(\ref{shhole}) (thin lines). Furthermore, as shown in Fig.
\ref{fig3}(b) for $\omega_0=0.05\Delta$ and $1/\tau\Delta=0.005$,
for $\omega \gtrsim\omega_0$ we find a weak deviation from
Eq.(\ref{shhole}) due solely to the imaginary part of the
self-energy, in contrast to Fig. \ref{fig2}(b) where the
spin-vertex corrections have a much stronger effect.

Before concluding, it is worth discussing how our results can be
obtained experimentally. In particular for the 2D linear Rashba
model we can make use of the equivalence between $\sigma_{\rm
sH}(\omega)$ and the longitudinal in-plane spin susceptibility
$\chi_\|(\omega)$ \cite{dimi}, and directly relate the poles of
$\sigma_{\rm sH}(\omega)$ with the time evolution of the spin
polarization $S_y(t)$ \cite{loss}, which can be measured by
various techniques \cite{fabian}. We find thus from
Eq.(\ref{sigma2}) that for $\tau\Delta\gg 1$, $S_y(t)$ is a
function oscillating with frequency $\Delta^*$, Eq.(\ref{reno}),
damped by an exponential decay with rate
$1/\tau_s=1/[2\tau(1+\lambda/2)]$. On the contrary, the decay
rate in the $\tau\Delta\ll 1$ limit is independent of the el-ph
interaction at $T=0$, and reduces to the Dyakonov-Perel value
$1/\tau_s=\Delta^2\tau/2$ \cite{fabian}.


FM: The hospitality of the Department of Condensed Matter Physics
at the University of Geneva is greatly appreciated. This work was
supported in part by the Natural Sciences and Engineering
Research Council of Canada (NSERC), by ICORE (Alberta), by the
Canadian Institute for Advanced Research (CIAR), and by the
University of Geneva.

\end{document}